\documentclass[useAMS,usenatbib, usegraphicx]{mn2e}
\usepackage{fancyhdr}
\usepackage{makeidx}
\usepackage{lscape}
\usepackage{wasysym}
\usepackage{longtable}
\usepackage{aas_macros}
\usepackage{txfonts}
\usepackage{times}
\usepackage{graphicx}
\usepackage{color}
\usepackage{ulem}

\begin{document}

\title[A deeper look at the GD1 stream: density variations and wiggles]{A deeper look at the GD1 stream: density variations and wiggles}
\author[T.J.L. de Boer et al.]{T.J.L. de Boer$^{1,2}$\thanks{E-mail:
t.deboer@surrey.ac.uk}, V. Belokurov$^{2,3}$, S. E. Koposov$^{2,4}$, L. Ferrarese$^{5}$,D. Erkal$^{1,2}$, P. C\^ot\'e$^{5}$ \newauthor and J.F. Navarro$^{6}$\\
$^{1}$ Department of Physics, University of Surrey, Guildford, GU2 7XH, UK\\
$^{2}$ Institute of Astronomy, University of Cambridge, Madingley Road, Cambridge, CB3 0HA, UK\\
$^{3}$Center for Computational Astrophysics, Flatiron Institute, 162 5th Avenue, New York, NY 10010, USA\\
$^{4}$ McWilliams Center for Cosmology, Department of Physics, Carnegie Mellon University, 5000 Forbes Avenue, Pittsburgh, PA 15213, USAs \\
$^{5}$ National Research Council of Canada, 5071 West Saanich Road, Victoria, BC, V9E 2E7, Canada \\
$^{6}$ Department of Physics and Astronomy, University of Victoria, Victoria, BC, Canada V8P 5C2
}
\date{Received ...; accepted ...}

\pagerange{\pageref{firstpage}--\pageref{lastpage}} \pubyear{2017}

\maketitle

\begin{abstract}
Using deep photometric data from CFHT/Megacam, we study the morphology and density of the GD-1 stream, one of the longest and coldest stellar streams in the Milky Way. Our deep data recovers the lower main sequence of the stream with unprecedented quality, clearly separating it from Milky Way foreground and background stars. An analysis of the distance to different parts of the stream shows that GD-1 lies at a heliocentric distance between 8 and 10 kpc, with only a shallow gradient across 45 deg on the sky. Matched filter maps of the stream density show clear density variations, such as deviations from a single orbital track and tentative evidence for stream fanning. We also detect a clear under-density in the middle of the stream track at $\varphi_{1}$=-45 deg surrounded by overdense stream segments on either side. This location is a promising candidate for the elusive missing progenitor of the GD-1 stream. We conclude that the GD-1 stream has clearly been disturbed by interactions with the Milky Way disk or other sub-halos.
\end{abstract}


\label{firstpage}

\section{Introduction}
\label{introduction}
Stellar streams provide a dramatic confirmation of the hierarchical galaxy formation scenario, showing us that large galaxies accrete smaller stellar systems \citep{Lynden-Bell95}. The streams are the relics of this formation process, consisting of stars stripping from the infalling host cluster or galaxy \citep{Newberg16}. Large scale, homogeneous sky surveys such as SDSS, 2MASS, Pan-STARRS, ATLAS, and DES \citep{Skrutskie06,Ahn14,Shanks15,Chambers16} have been successfully mined for stream features, resulting in nearly a dozen streams to date \citep{DES_streams}. These streams show a wide variety of morphologies, ranging from the wide and long Sagittarius stream to the very short Ophiuchus stream \citep{Belokurov06,Bernard14}. The orbits of streams have been used to place some of the strongest constraints on the mass distribution of the Milky Way (MW) to date \citep[e.g.,][]{Koposov10,Law10a,Gibbons14,Bowden15,Kuepper15,Bovy16}. Streams also represent one of the very few ways of probing the existence of dark matter~(DM) subhalos \citep[e.g.][]{Ibata02,Johnston02,Carlberg09,Yoon11,Carlberg12,Erkal15}, which are not expected to form any stars \citep{Ikeuchi86,Rees86}. Analysis of deep data for the Pal 5 stream has indeed shown tentative evidence of a disturbance by low mass subhaloes \citep{Bovy17,Erkal17}.

The GD-1 tidal stream was first discovered by \citet{Grillmair061} in the Sloan Digital Sky Survey (SDSS) as a 63 degree long trail of stars extending from the constellation of Ursa Major to Cancer. The stream is very thin on the sky ~($\sigma\approx$0.2 deg), indicating that it most likely originated in a globular cluster (GC) stripping event~\citep{Koposov10}. GD-1 is among the most suitable streams for the study of the lumpiness of DM, satisfying several necessary conditions. First, the stream has a small initial phase-space volume, facilitating a clean separation of the velocity kicks induced by dark substructures from the secular dynamics of the progenitor. Moreover, the stream is long enough to have experienced many possible interactions, and has a high surface brightness which allows us to measure the stream density with high accuracy \citep{Erkal16}. This makes it easier to detect density perturbations along the stream. 

While the theory behind stream gap formation is well developed \citep{Yoon11,Carlberg12,Erkal15,Sanders16}, determining whether an observed gap is the result of a dark subhalo passage is more difficult. For many streams (including the Pal 5 stream) gaps may results from an interaction with a giant molecular cloud \citep{Amorisco16} or the MW bar \citep{Erkal17, Pearson17}. However, the derived orbits of the GD-1 stream show that it is moving in a retrograde sense with respect to the disk with a perigalacticon of about 14 kpc and apogalacticon of 26-29 kpc \citep{Koposov10,Bowden15}. This would imply that interactions with disk sub-strucure are less likely, and thus make it an ideal candidate for the study of gaps induced by dark subhalos.

To date, the stellar content of the stream has not yet been studied in depth, due to the relatively shallow photometric data and small numbers of spectroscopic members. Based on isochrone fitting of SDSS photometry, the stream is best matched to a relatively metal-rich population with [Fe/H]=-1.4 and an age of $\approx$9 Gyr~\citep{Koposov10}. However, based on spectroscopic measurements from the SDSS SEGUE survey~\citep{Yanny091} a more metal-poor population with [Fe/H]=-1.9 and ancient age is favoured~\citep{Willett09}. The stream does not have any apparent progenitor, but its low spectroscopic metallicity and small angular width suggest that a GC origin is more appropriate than a dwarf galaxy. Although it is possible that the progenitor may be located outside the currently studied stream footprint, no known star cluster or galaxy has been found to match the orbital properties of the stream. It is also possible the parent satellite has by now fully dissolved, leaving no detectable remnant behind.

Studies using SDSS photometry have indicated that there may be density variations along the GD-1 stream. Analysis by~\citet{Koposov10} shows the presence of several under-densities of varyious sizes, which do not correlate with CCD gaps in the Sloan camera, fluctuations in dust extinction or other sources of spurious detections. More recently,~\citet{Carlberg13} investigated the same SDSS data for the presence of stream gaps. A total of 8 tentative gaps are detected along the stream, albeit with limited confidence. The number of large gaps is in good agreement with predictions for DM sub-halo encounters with cold star streams, while the number of small gaps is well below predictions~\citep{Carlberg13}. The mechanism responsible for these gaps is not clear, in part due to the lack of a progenitor~\citep{Kuepper08,Kuepper12}. More recently, \cite{Erkal16} predicted the number of gaps from subhaloes expected in GD-1 and argued that there should be no small gaps but rather roughly 1 gap with a size on the order of $\sim 6.5$ degrees if GD-1 were on a circular orbit. Since GD-1 is currently close to pericenter \citep{Koposov10,Bowden15}, the gap should be larger than this on average.

In this work, we investigate the spatial distribution of GD-1 stream stars, using deep, wide-field CFHT data. This new data goes $\approx$2 magnitudes deeper than the shallow SDSS photometry and reaches the level of faint Main Sequence stars across a 45 degree stretch of the stream. The greater depth affords a cleaner separation of the GD-1 features from those of the Milky Way stellar halo, allowing us to determine more accurate distances than possible from SDSS photometry. We study the evolution of the stream centroid on the sky and search for variations such as wiggles around the mean track, which are indicative of perturbations due to tidal encounters \citep{Carlberg12,Erkal15,Erkal15b}. Finally, we study the stream's density profile on large scales as well as small-scale density variations with greater spatial resolution than previously possible, and look for under-densities and gaps. This provides a new, clearer view of the GD-1 stream.

This paper is organised as follows: in Section~\ref{data} we present the CFHT photometric dataset used, the data reduction procedure and discuss the resulting catalog. In Section~\ref{trackdist} we determine the stream track and the distance to different parts of the stream. This is followed in Section~\ref{GD1density} by matched filter maps of GD-1 and an analysis of stellar density variations. Finally, Section~\ref{conclusions} discusses the results and their implications for the lumpiness of DM.
\section{Data}\label{data}
Deep optical photometry of the GD-1 stream in the g and r filters was obtained using the Megacam camera on CFHT as part of observing proposals 15AC13 and 16AC03 (PI: de Boer). Observations consisted of 4 dithered exposures of 50s each, in both g and r bands, under dark conditions and seeing less than 0.8 arcsec. Pointings were overlapped and dithered to fill CCD and tiling gaps and obtain the most homogeneous spatial coverage. The images were astrometrically calibrated against SDSS and subsequently swarped to a common grid with a maximum number of 4 stacked images, to achieve a homogeneous data depth~\citep{Bertin02}. Following this, the images were reduced with the help of SExtractor and PSFex~\citep{Bertin96}. The resulting catalogues were photometrically calibrated using SDSS photometry and corrected for extinction using dust maps from~\citet{Schlegel98} with coefficients from \citet{Schlafly11}, on a star by star basis.

\begin{figure}
\centering
\includegraphics[angle=0, width=0.495\textwidth]{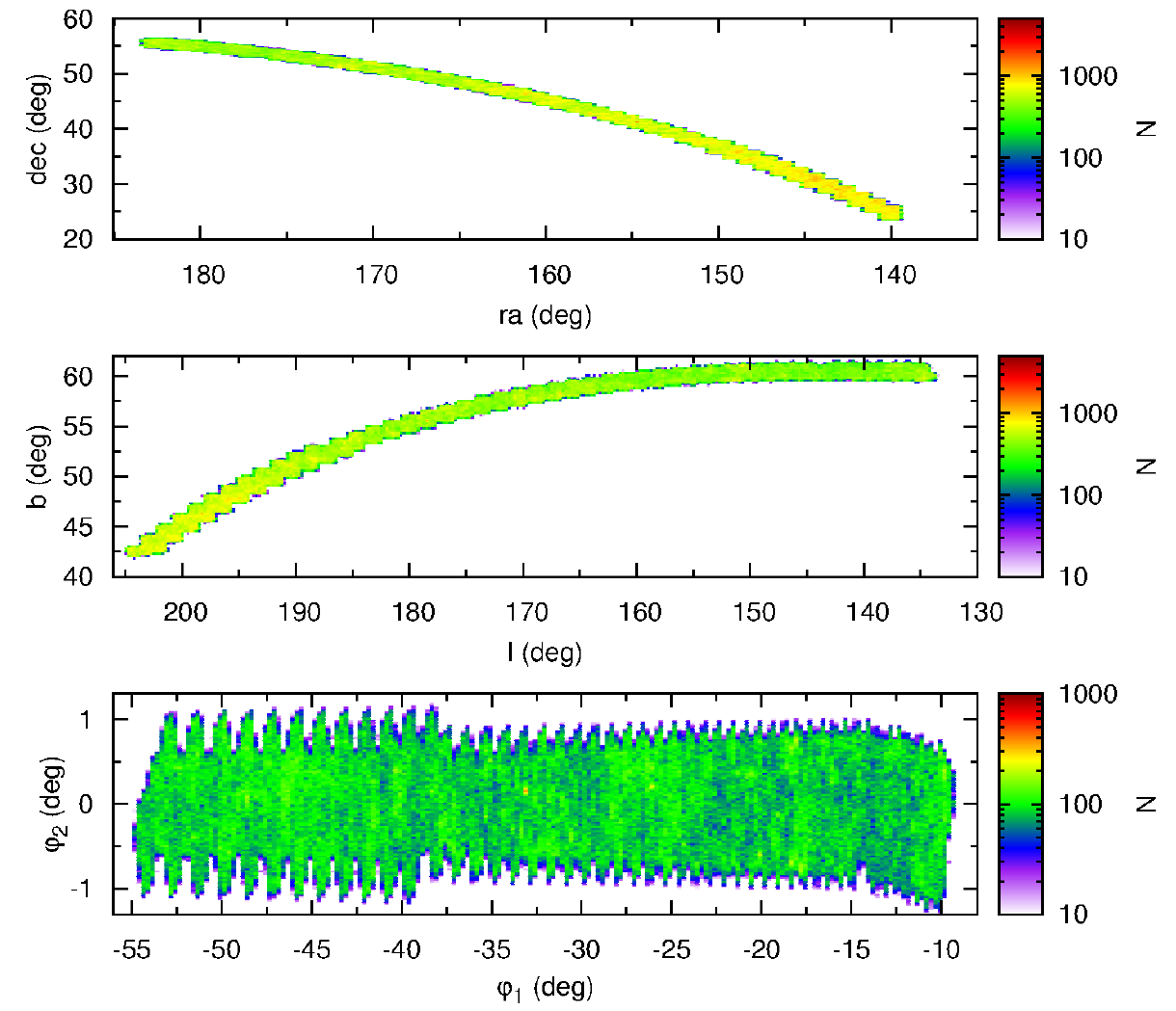}
\caption{Spatial coverage of all stars in the CFHT/Megacam photometry of the GD-1 stream, in equatorial coordinates ra,dec (top), Galactic coordinates l,b (middle) and in coordinates aligned with the stream~(bottom). The colour indicates the density of stars across the footprint. \label{GD1_spatial}}
\end{figure}

The spatial coverage of the deep photometry is shown in Figure~\ref{GD1_spatial} in equatorial, Galactic and rotated-spherical heliocentric great circle coordinates aligned with the stream~\citep{Koposov10}. The observations resulted in a total coverage of 45 degrees along the stream and out to  at least 4 times the width of the stream. No obvious signs of pointing or tiling marks are visible in the bottom panel of Figure~\ref{GD1_spatial}, showing that our catalogue is flat and homogeneous across the footprint.

\begin{figure}
\centering
\includegraphics[angle=0, width=0.495\textwidth]{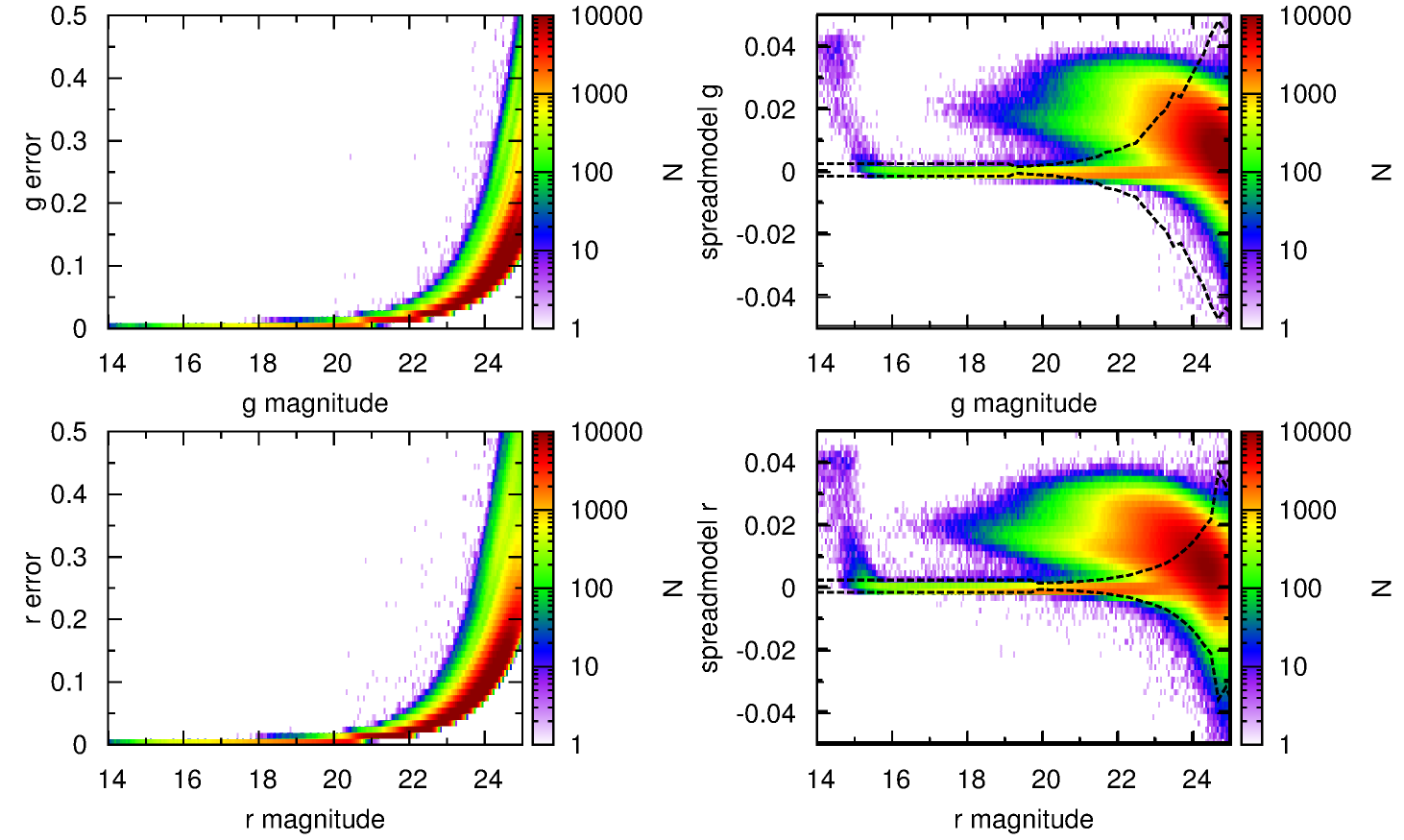}
\caption{Data quality overview showing the photometric errors as a function of g and r filters in the left panels and the $\texttt{SPREAD\_MODEL}$ parameter in the right panels. The black dashed lines in the right panels indicate the $\texttt{SPREAD\_MODEL}$ selection adopted to remove unresolved galaxies from the catalog. \label{GD1_errs}}
\end{figure}

Figure~\ref{GD1_errs} displays the photometric error as a function of magnitude for each filter for all stars in the catalogue. The photometric error profile is narrow and well defined, showing no hints of pointings with anomalous depth. The right panels of Figure~\ref{GD1_errs} show the SExtractor $\texttt{SPREAD\_MODEL}$ parameter as function of magnitude. Our final sample of stars are selected using a cut at $|\texttt{SPREAD\_MODEL}|<$0.002 +$\texttt{SPREADERR\_MODEL}$ to discriminate between likely stars and unresolved galaxies.

\begin{figure}
\centering
\includegraphics[angle=0, width=0.495\textwidth]{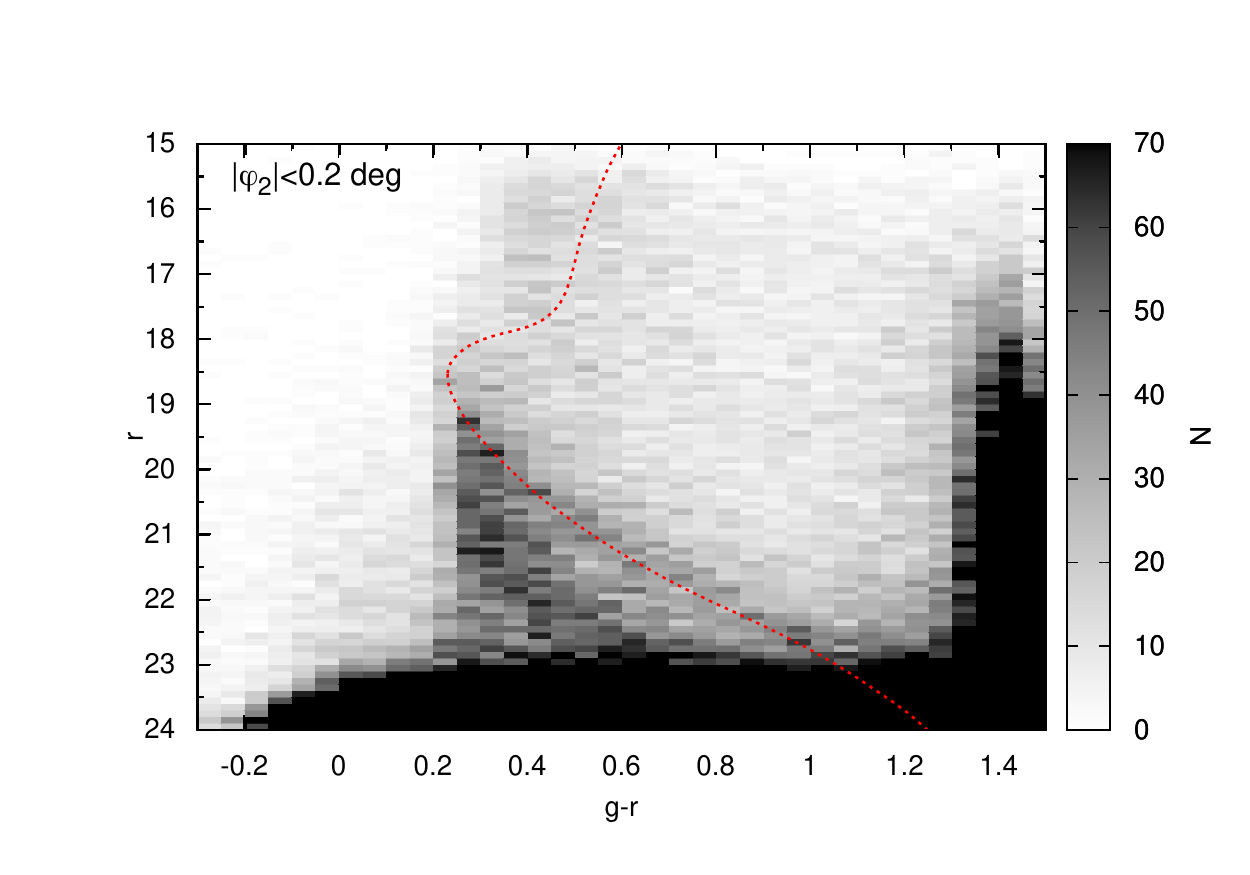}
\includegraphics[angle=0, width=0.495\textwidth]{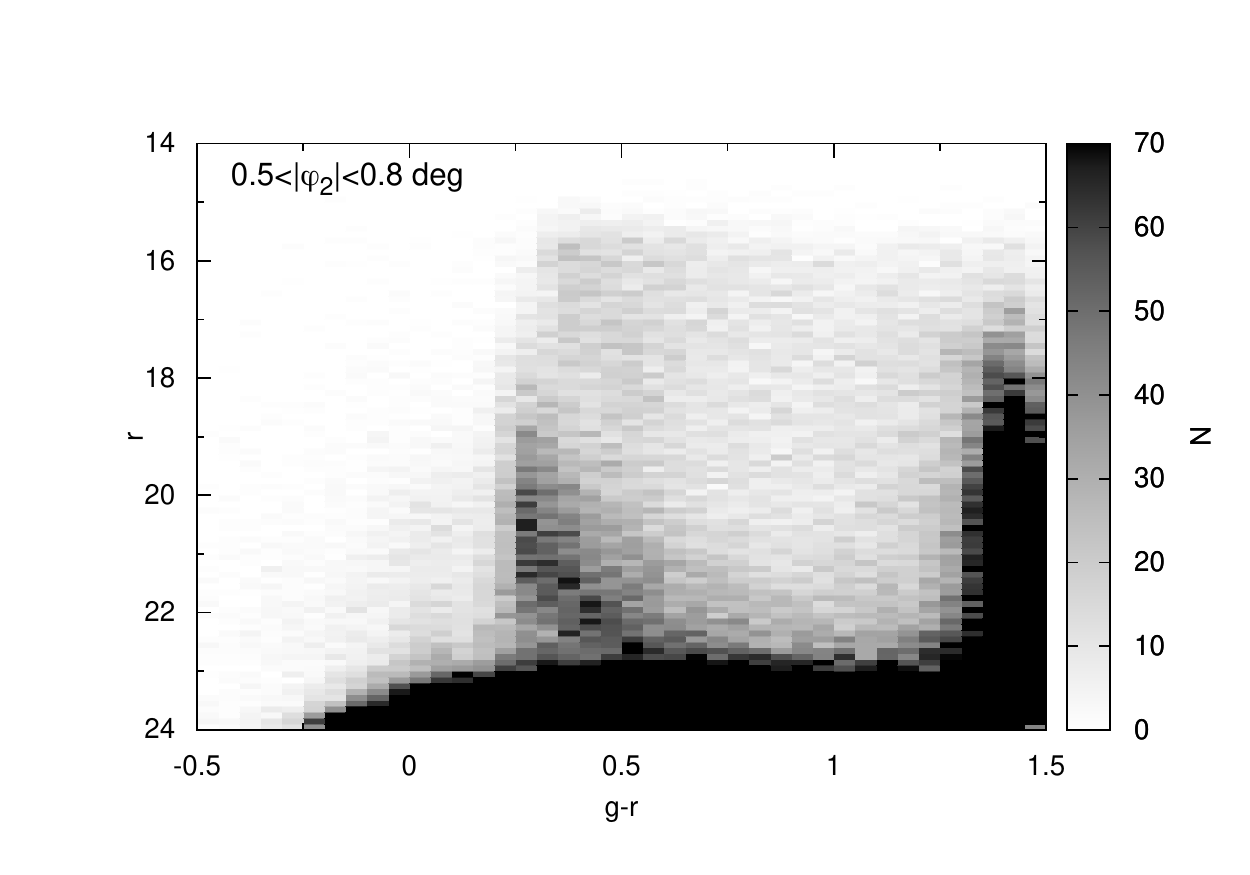}
\caption{The colour-magnitude diagram of our data, overlaid with a reference isochrone for GD-1 with parameters ([Fe/H]=$-$1.9, 12 Gyr) at a distance of 9 kpc. The top panel shows the observed Hess diagram of all stars in the stream with $|\varphi_{2}|<$0.2 degrees, while the bottom panel shows the Hess diagram for the background using 0.5$<|\varphi_{2}|<$0.8 degrees. The lower main sequence and turn-off of the GD-1 population is clearly visible, as traced by the reference isochrone. \label{GD1_CMDs}}
\end{figure}

Figure~\ref{GD1_CMDs} shows the (r,g$-$r) CMD of all stellar objects in our data following the centre of the stream coordinate frame with $|\varphi_{2}|<$0.2 degrees. The bottom panel also shows the CMD of our off-stream background region with 0.5$<|\varphi_{2}|<$0.8 degrees, which is free from stream stars. The CMDs display all typical features commonly associated with the MW foreground (including the red disk MS stars at g$-$r=1.5 and faint blue halo stars at g$-$r=0.3). However, the CMD of the stream region also displays a clear diagonal overdense sequence of GD-1 stream stars, extending from g$-$r=0.3, r=19 to g$-$r=1.1, r=23 which corresponds to the lower MS and turn-off of the stream. 
This GD-1 sequence clearly stands out from the underlying MW halo contamination, and is well traced by the reference isochrone for GD-1 populations. We adopt a reference isochrone with [Fe/H]=$-$1.9, an age of 12 Gyr, and a distance of 9 kpc is from the Dartmouth isochrone library~\citep{DartmouthI}.

The CMD shows the presence of overdense features on the blue side of the CMD with g$-$r$\approx$0 fainter than the GD-1 MSTO. They have colours reminiscent of Blue Horizontal Branch or Blue Straggler Stars, but could also constitute a more metal-poor MS turn-off at larger distance. Therefore, we select all stars with -0.1$<$g-r$<$0.5 and 22$<$r$<$23 and show their spatial density map in Figure~\ref{GD1_density_faint}. Two clear sequences are visible in the density map, perpendicular to the direction of the GD-1 stream. Comparison to the overlaid stream tracks shows that these features correspond to the Sgr stream and Orphan stream, both of which are further away and more metal-poor than the GD-1 stream.

\section{Distance and stream track}\label{trackdist}
With calibrated data in hand, we turn to the determination of the stream parameters, such as distance and position on the sky. Due to the greater depth and increased accuracy of the CFHT data over SDSS catalogues, we can make use of the faint MS to determine the distance to the GD-1 stream. Therefore, we select stars with main sequence colours (0.5$<$g-r$<$0.9) and magnitudes corresponding to dwarfs (19$<$r$<$23) and compare their magnitude to those of the GD-1 reference isochrone. For each star, we compute the difference between its r-band magnitude and the magnitude expected for the isochrone. These differences are then summed up to create a distance modulus histogram for each bin of $\varphi_{1}$ considered. 

To bring out the signal of the GD-1 stream, we correct for the effects of MW contamination using the Galaxia model of the MW~\citep{Sharma11}. A Galaxia realisation is generated in small bins of Galactic longitude and latitude along the stream footprint and convolved with the photometric scatter due to magnitude dependent errors shown in Figure~\ref{GD1_errs}. We then compute the same distance modulus offsets for the Galaxia realisation as for the observed CMD and subtract it from the on-stream results.

\begin{figure}
\centering
\includegraphics[angle=0, width=0.495\textwidth]{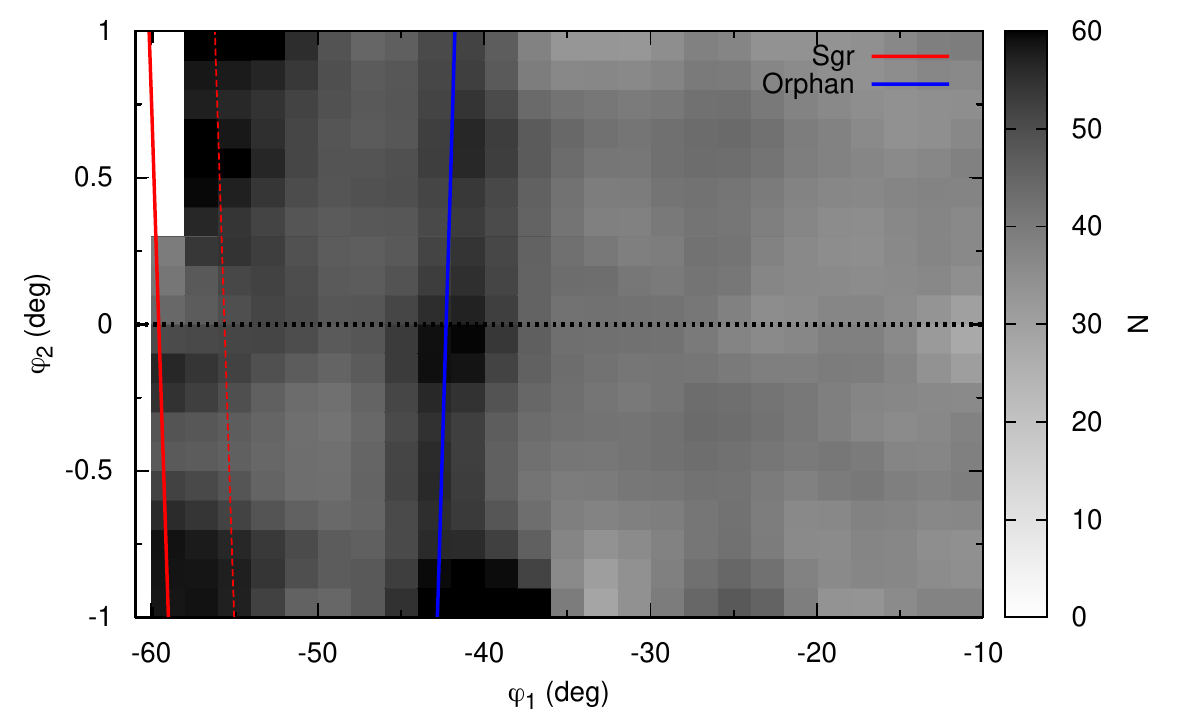}
\caption{The spatial density of faint, blue stars (-0.1$<$g-r$<$0.5 and 22$<$r$<$23) across the GD-1 footprint. Lines indicate the trajectory of known streams in this part of the sky, such as the Sagittarius bright stream~(red lines) and the Orphan stream in blue~\citep{Belokurov072, Koposov12}. \label{GD1_density_faint}}
\end{figure}

\begin{figure}
\centering
\includegraphics[angle=0, width=0.495\textwidth]{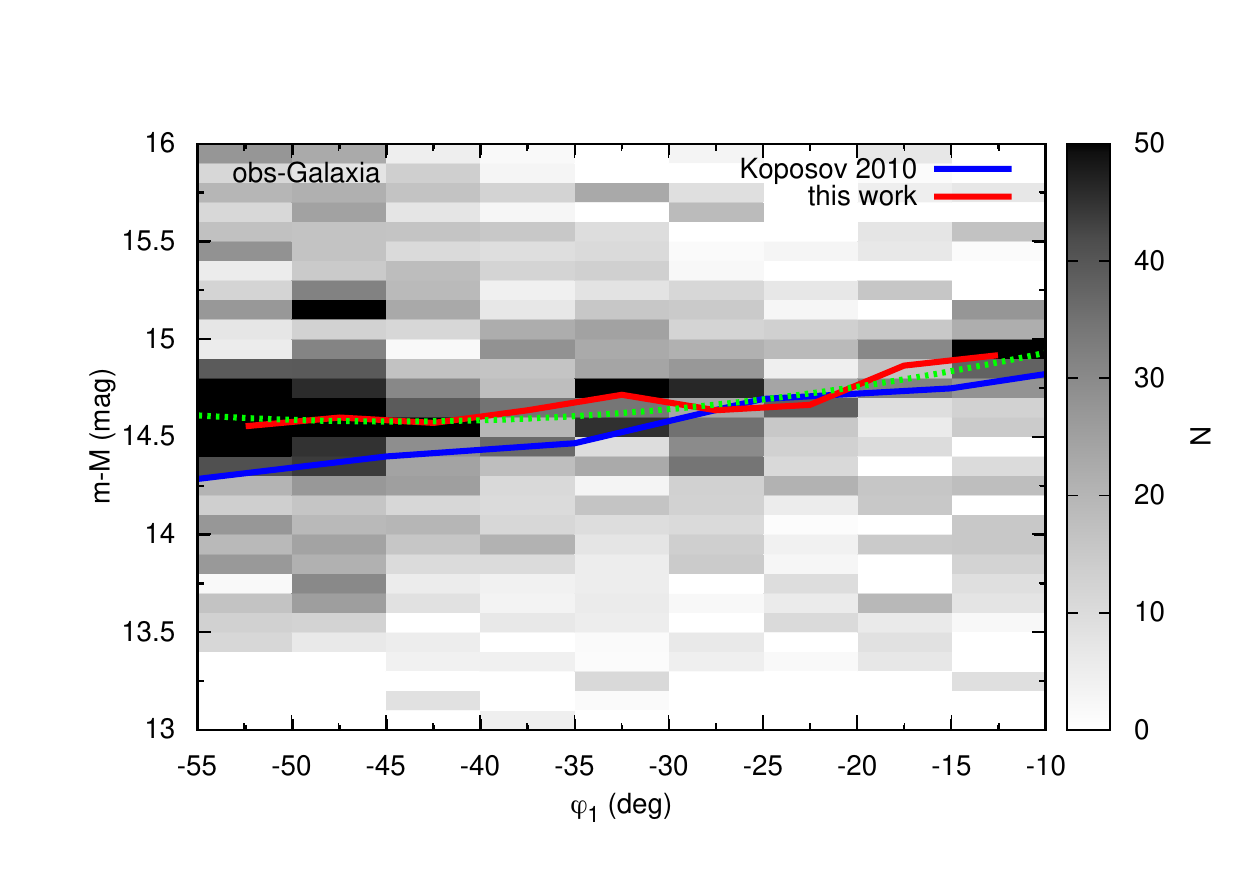}
\caption{MS distance modulus as a function of $\varphi_{1}$ for the GD-1 stream. The figure shows the density of stars in magnitudes offset from the MS between 0.5$<$g-r$<$0.9 and r$>$19, after MW correction. The red line indicates the location of the best Gaussian fit to each bin of $\varphi_{1}$, while the blue line indicates the distance determination from~\citep{Koposov10}. The green (dotted) line shows a best fit second order polynomial to the distances, described by the function f($\varphi_{1}$)=14.58 +(2.923e-4*($\varphi_{1}$+44.66)$^{2}$). \label{GD1_distfit}}
\end{figure}

Figure~\ref{GD1_distfit} displays the MS distance modulus for the GD-1 stream, as determined for our stellar sample after MW correction. A clear sequence is visible, for distance modulus of $\approx$14.7 or $\approx$9 kpc. To determine the heliocentric distance at each position in the stream, we fit a Gaussian to the density histogram in each $\varphi_{1}$ along with a second order polynomial background and determine the centre and standard deviation. The red line in Figure~\ref{GD1_distfit} indicates the determined distance to GD-1, while the blue line indicates the results from~\citet{Koposov10}. Values for the Gaussian centre and standard deviation are also given (along with their uncertainties) in Table~\ref{distvals}. Finally, the green (dotted) line shows a second order polynomial fit to the distances, as described by the function f($\varphi_{1}$)=14.58 +(2.923e-4*($\varphi_{1}$+44.66)$^{2}$). This function will be subsequently used to correct for the distance to GD-1. 
Figure~\ref{GD1_distfit} indicates that GD-1 shows a gradually increasing distance as $\varphi_{1}$ increases, ranging from $\approx$8 to 10 kpc. Compared to the results of~\citet{Koposov10}, our data indicate the stream is on average slightly more distant but in good agreement at higher $\varphi_{1}$. Our new results show less variation in the distance to GD-1 in this part of its footprint, with a change of only 1.5 kpc over 45 degrees. The derived results correspond to a range in galactocentric distances between 13 and 15 kpc, which is consistent with the perigalacticon distance of best-fit GD1 orbits \citep{Carlberg13}.

With the distance to the stream in place, we next determine the track of the stream on the sky. To determine the stream track, we first perform a matched filtering of the data, assuming that the GD-1's CMD is well described by an isochrone with [Fe/H]=$-$1.9, 12 Gyr and distances from Table~\ref{distvals}. Similar to~\citet{Erkal17}, we forgo using a classical matched filter technique (in which weighted data is no longer Poisson distributed) and instead adopt a boolean mask procedure. 
For the signal mask we populate the [Fe/H]=$-$1.4, 9 Gyr isochrone using a Kroupa IMF~\citep{Kroupa01} and convolve the synthetic stars with the photometric errors as function of magnitude extracted from the observed data. To avoid a filter that is arbitrarily thin at bright magnitudes (where photometric errors are small), a minimum filter width of 0.05 is adopted. For the background filter, we generate an oversampled Galaxia MW model for each pixel (to ensure enough stars populate the CMD) and once again convolve stars with the observed photometric errors. Both masks are normalised, and the fraction P$_{str}$(g-r,r)/P$_{bg}$(g-r,r) is used to select CMD pixels below a certain threshold. For each pixel, the threshold value is chosen by looping through the possible values and finding the one which maximises the signal-to-noise. We then assign a weight of zero to values of P$_{str}$(g-r,r)/P$_{bg}$(g-r,r) lower than the threshold and a weight of 1 for value higher than the threshold.

From the analysis of the CMD in Figure~\ref{GD1_CMDs} we have shown signs of the Orphan stream population in our data, partially overlapping with the main sequence of GD-1. The density of Orphan stars intersecting the GD1 main sequence is low, given its much greater distance. Nonetheless, given the position of the Orphan stream in our footprint (see Figure~\ref{GD1_density_faint}), contamination might influence the GD-1 stream track at low $\varphi_{1}$. Therefore, we include an extra background filter during the matched filter process, based on the Orphan stream stellar populations and distance. This ensures that CMD bin coincident with the Orphan stream stellar population locus are penalised in the filtering, thereby avoiding Orphan stream contamination.

\begin{table}
\caption[]{Distance modulus and heliocentric distance to the GD-1 stream for bins of $\varphi_{1}$ position.}
\begin{center}

\begin{tabular}{ccccc}
\hline\hline
$\varphi_{1}$ &  m-M & $\sigma_{m-M}$ & distance & $\sigma_{distance}$  \\
(deg) & (mag) & (mag) & (kpc) & (kpc)  \\
\hline
-52.5 & 14.56$\pm$0.01 & 0.12$\pm$0.01 & 8.15$\pm$0.04 & 0.45$\pm$0.01 \\ 
-47.5 & 14.60$\pm$0.04 & 0.11$\pm$0.04 & 8.32$\pm$0.15 & 0.42$\pm$0.14 \\
-42.5 & 14.57$\pm$0.05 & 0.09$\pm$0.05 & 8.21$\pm$0.19 & 0.34$\pm$0.18 \\
-37.5 & 14.64$\pm$0.09 & 0.20$\pm$0.01 & 8.46$\pm$0.35 & 0.78$\pm$0.03 \\
-32.5 & 14.72$\pm$0.05 & 0.14$\pm$0.06 & 8.78$\pm$0.20 & 0.57$\pm$0.22 \\
-27.5 & 14.64$\pm$0.06 & 0.20$\pm$0.01 & 8.46$\pm$0.23 & 0.78$\pm$0.03 \\
-22.5 & 14.67$\pm$0.10 & 0.20$\pm$0.11 & 8.57$\pm$0.40 & 0.79$\pm$0.43 \\
-17.5 & 14.87$\pm$0.07 & 0.10$\pm$0.07 & 9.40$\pm$0.30 & 0.42$\pm$0.31 \\
-12.5 & 14.92$\pm$0.04 & 0.10$\pm$0.04 & 9.63$\pm$0.18 & 0.44$\pm$0.17 \\
\hline 
\end{tabular}
\end{center}
\label{distvals}
\end{table}

\begin{table}
\caption[]{Position and width of the GD-1 stream, as shown in Figure~\ref{GD1_trackfit}.}
\begin{center}

\begin{tabular}{ccc}
\hline\hline
$\varphi_{1}$ &  $\varphi_{2}$ & $\sigma_{\varphi_{2}}$  \\
(deg) & (deg) & (deg)  \\
\hline
 -53 & -0.28$\pm$0.12 & 0.06$\pm$0.13 \\
 -51 & -0.09$\pm$0.09 & 0.35$\pm$0.25 \\
 -49 & -0.03$\pm$0.10 & 0.35$\pm$0.23 \\
 -47 & -0.01$\pm$0.09 & 0.35$\pm$0.23 \\
 -45 &  0.01$\pm$0.10 & 0.35$\pm$0.20 \\
 -43 & -0.05$\pm$0.17 & 0.16$\pm$0.29 \\
 -41 & -0.05$\pm$0.19 & 0.22$\pm$0.22 \\
 -39 & -0.02$\pm$0.16 & 0.10$\pm$0.18 \\
 -37 &  0.00$\pm$0.11 & 0.10$\pm$0.12 \\
 -35 &  0.04$\pm$0.10 & 0.13$\pm$0.11 \\
 -33 &  0.05$\pm$0.09 & 0.15$\pm$0.11 \\
 -31 &  0.05$\pm$0.08 & 0.15$\pm$0.10 \\
 -29 &  0.04$\pm$0.08 & 0.17$\pm$0.10 \\
 -27 &  0.04$\pm$0.09 & 0.19$\pm$0.15 \\
 -25 &  0.06$\pm$0.10 & 0.18$\pm$0.15 \\
 -23 &  0.08$\pm$0.10 & 0.16$\pm$0.14 \\
 -21 &  0.08$\pm$0.14 & 0.24$\pm$0.29 \\
 -19 &  0.09$\pm$0.19 & 0.35$\pm$0.88 \\
 -17 & -0.04$\pm$0.11 & 0.14$\pm$0.14 \\
 -15 & -0.08$\pm$0.08 & 0.13$\pm$0.10 \\
 -13 & -0.14$\pm$0.08 & 0.13$\pm$0.10 \\
 -11 & -0.21$\pm$0.18 & 0.23$\pm$0.39 \\
\hline 
\end{tabular}
\end{center}
\label{GD1_trackvals}
\end{table}

\begin{figure*}
\centering
\includegraphics[angle=0, width=0.95\textwidth]{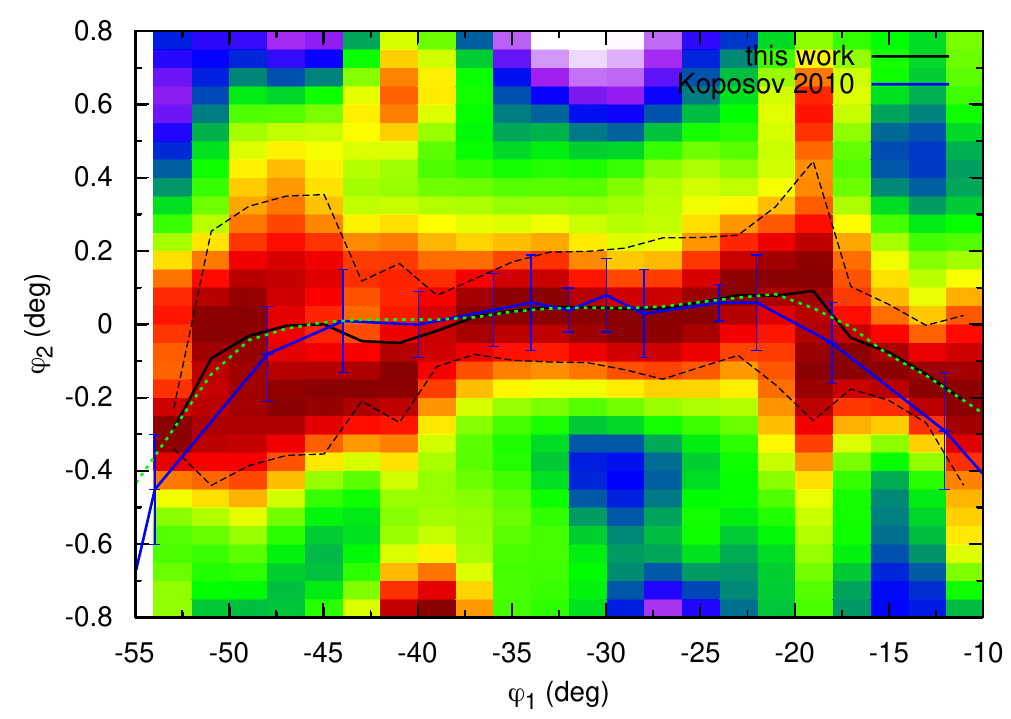}
\caption{Matched filter map of the GD-1 stream using spatial bins of 2$\times$0.05 deg. The map has been normalised in each $\varphi_{1}$ column. The black solid and dashed lines indicate the best-fit Gaussian location and standard deviation of the stream track at each $\varphi_{1}$ position. The blue line and error bars indicate the stream positions from~\citet{Koposov10}. Finally, the green line shows the smoothed track after applying a boxcar filter. \label{GD1_trackfit}}
\end{figure*}

Figure~\ref{GD1_trackfit} shows the matched filter map of GD-1 using rough spatial bins of 2$\times$0.05 deg. To bring out low contrast stream features in more detail, the map is convolved using a Gaussian kernel of 2$\times$0.05 deg, and normalised in each $\varphi_{1}$ column. A clear sequence is visible around $\varphi_{2}\approx$0 deg, which corresponds to the track of the GD-1 stream. For reference, the stream track of GD-1 derived in~\citet{Koposov10} is overlaid as the blue line with errorbars. The track of GD-1 extracted here is roughly consistent with the ~\citet{Koposov10} track, showing a distinct down-turn at both sides of our footprint. However, there are also deviations from the track, particularly on the left side, which corresponds to low Galactic latitude. To fit the GD-1 stream track in our new data, we fit a Gaussian model and background to each column of $\varphi_{1}$. The recovered stream track is shown as the solid black line in Figure~\ref{GD1_trackfit}, with the width indicated by dashed lines. The values for the track centre and width are also listed in Table~\ref{GD1_trackvals}. Figure~\ref{GD1_trackfit} shows the GD-1 stream is very narrow in the middle -40$<\varphi_{1}<$-25 deg stretch, with a total width of $\approx$0.3 deg. At low $\varphi_{1}$ values, the stream becomes much broader and appears to show some wiggles. 

The stream track is not smooth and shows some wiggles due to uncertainties in the track determination, which could introduce artificial wiggles in the stream density around the track. Therefore, we smooth the track using a simple boxcar filter, resulting in the green dotted line in Figure~\ref{GD1_trackfit}. This smooth track will subsequently be used to determine the density and variation of the GD-1 stream.

\begin{figure}
\centering
\includegraphics[angle=0, width=0.495\textwidth]{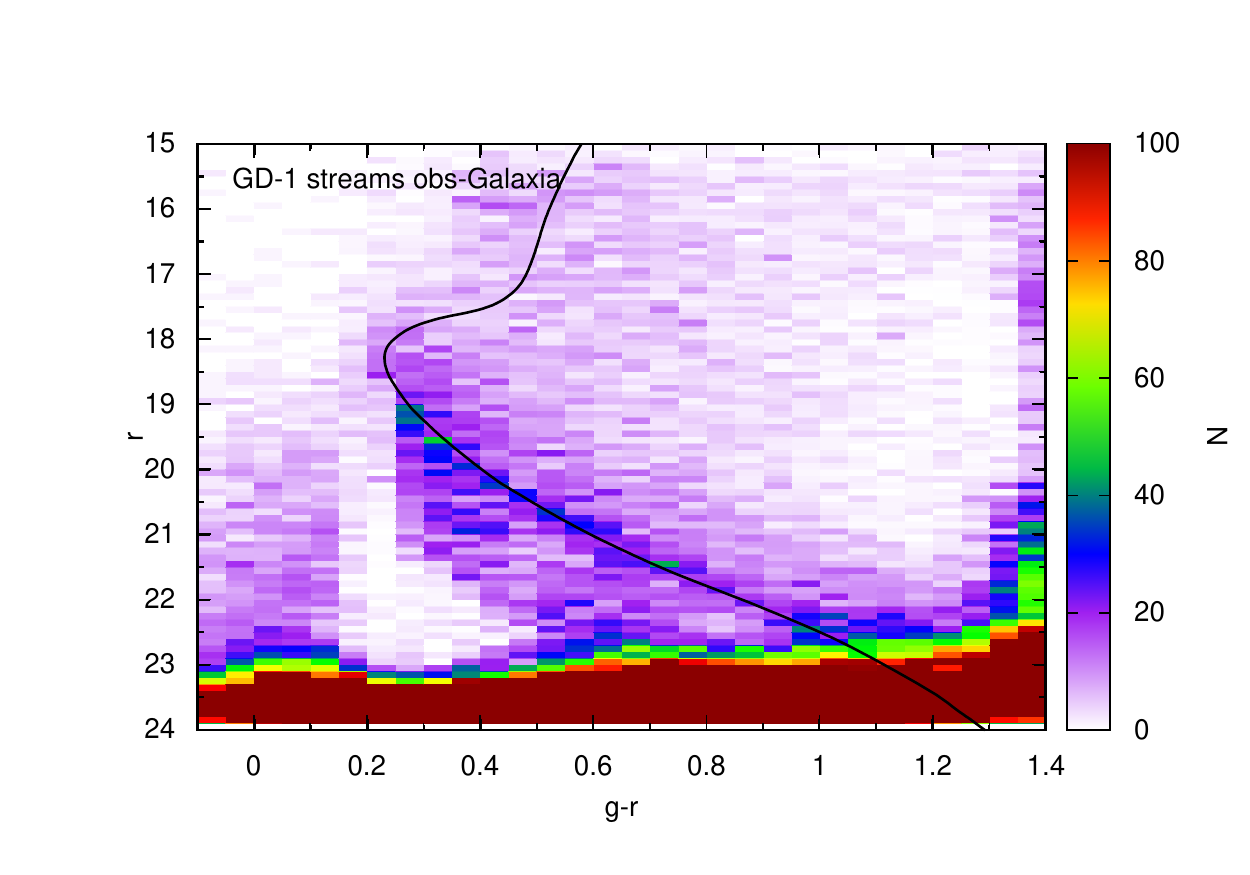}
\caption{The colour-magnitude diagram of the GD-1 stream, extracted around the stream track determined from Figure~\ref{GD1_trackfit} and corrected for distance variations. A reference isochrone for GD-1 is overlaid with parameters ([Fe/H]=$-$1.9, 12 Gyr). The top panel shows the observed Hess diagram of all stars, while the bottom panel shows the Hess diagram after correcting for MW contamination using the Galaxia model of the MW~\citep{Sharma11}. The lower main sequence and turn-off of the GD-1 population is clearly visible, as traced by the reference isochrone. \label{GD1_CMD_trackdistcorr}}
\end{figure}

To check the validity of the track and distance determination, we extract the CMD of GD-1 stars around the stream track, and correct for the distance variation, shifting stars to a common distance modulus of m-M=14.50. The g$-$r,r CMD after subtracting the Galaxia MW model realisation is shown in Figure~\ref{GD1_CMD_trackdistcorr}, along with the GD-1 reference isochrone overlaid. The GD-1 main sequence and turn-off are both clearly visible, along with the start of the sub-giant branch. The GD-1 main sequence appears much sharper in the corrected CMD, compared to Figure~\ref{GD1_CMDs}, indicating a good distance correction. The narrow width of the main sequence is obvious, ruling out a composite stellar population. This shows once again the GD-1 stream was likely formed from the stripping of a GC, instead of a dwarf galaxy.

\section{GD-1 stream density}\label{GD1density}
\begin{figure*}
\centering
\includegraphics[angle=0, width=0.9\textwidth]{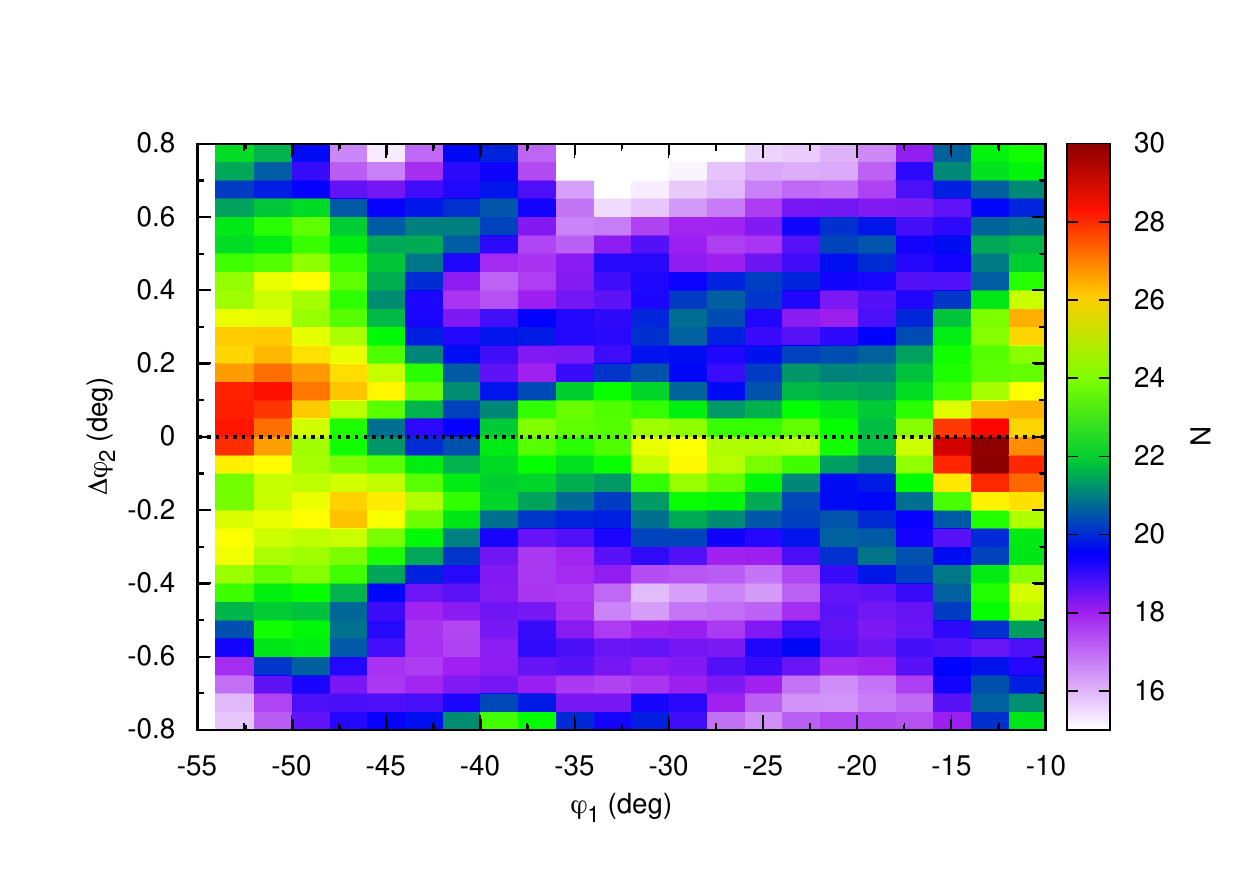}
\includegraphics[angle=0, width=0.9\textwidth]{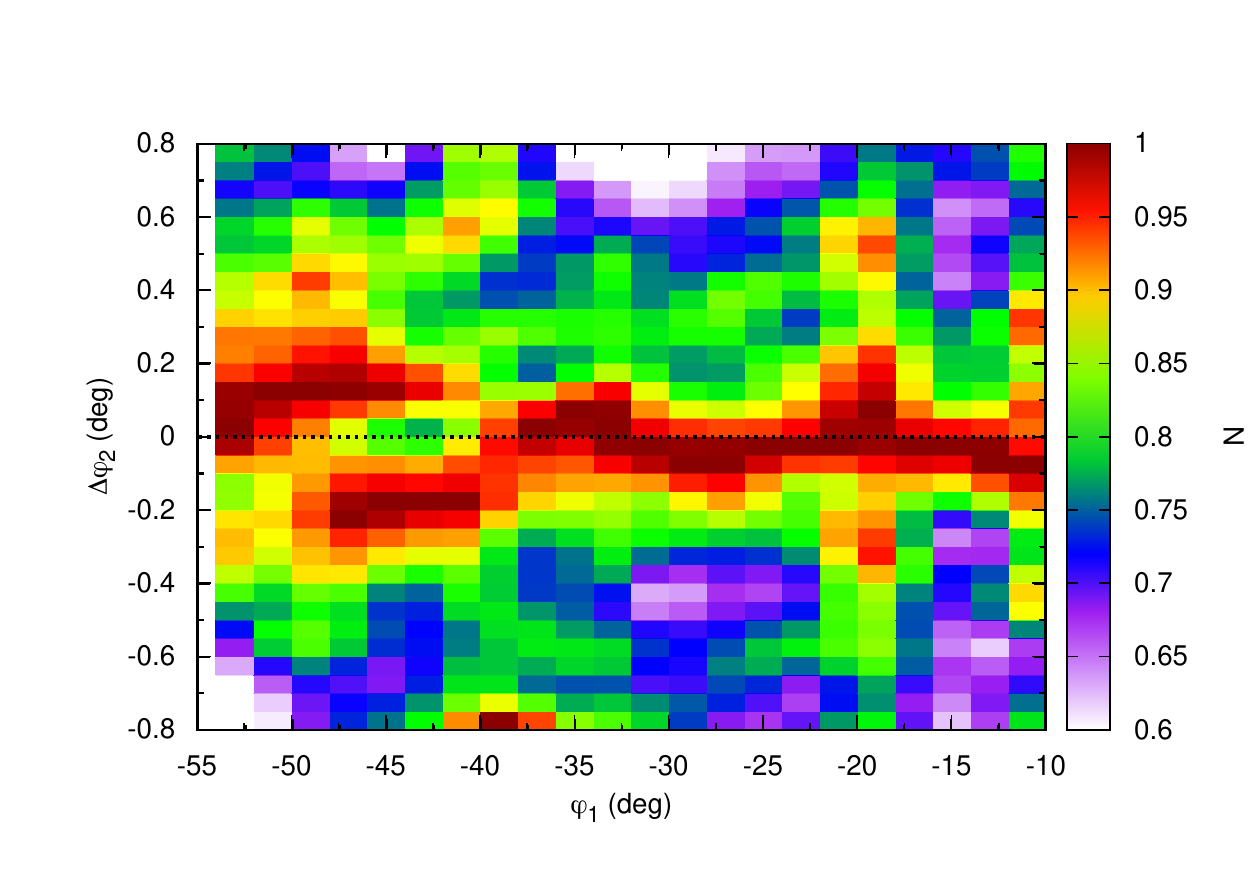}
\caption{Matched filter map of the GD-1 stream after stream track correction, using spatial bins of 2$\times$0.1 deg bins. The top panel shows a convolved matched filter density map, while the bottom panel shows a column normalised map to increase the stream track contrast. The maps have been convolved using a Gaussian kernel of 2$\times$0.1 deg. \label{GD1_MFmap}}
\end{figure*}

With the distance and stream track determined, we can now start constructing density maps of the GD-1 stream with greater resolution. Similar to Section~\ref{trackdist}, we once again perform a matched filter mapping of the data. However, this time, we offset the data in $\varphi_{2}$ using the stream track positions of Table~\ref{GD1_trackvals} to be able to see variations and wiggles of GD-1 around the track. The maps are constructed using spatial bins of 2$\times$0.1 deg to reveal the stream density in greater detail. Figure~\ref{GD1_MFmap} shows the matched filter map of the GD-1 stream, before (top panel) and after (bottom panel) column normalisation. In the top panel of Figure~\ref{GD1_MFmap}, we show the resulting matched filter density map of the GD-1 stream. To bring out the shape of the GD-1 stream track, we also show a column normalised version of the map in the bottom panel. This map is constructed by individually normalising the stream density in each column of $\varphi_{1}$. A column normalised map is often used to bring out connected low density features in the presence of varying background contamination, and will highlight the region of high density at each $\varphi_{1}$. 

As expected, the stream density largely follows the stream track determined in Section~\ref{trackdist}, with density peaking around $\varphi_{2}$=0. There are however strong density variations as function of $\varphi_{1}$, with an obvious lower density area around $\varphi_{1}\approx$-40 deg, and higher density regions at either end of our footprint. The column normalised map in Figure~\ref{GD1_MFmap} shows that GD-1 displays clear wiggles around the stream track. Apart from the large deviation at $\varphi_{1}$=-45 deg, the stream shows wiggles offset from the track by an amount smaller than its width of $\approx$0.1 deg. 

\begin{figure*}
\centering
\includegraphics[angle=0, width=0.495\textwidth]{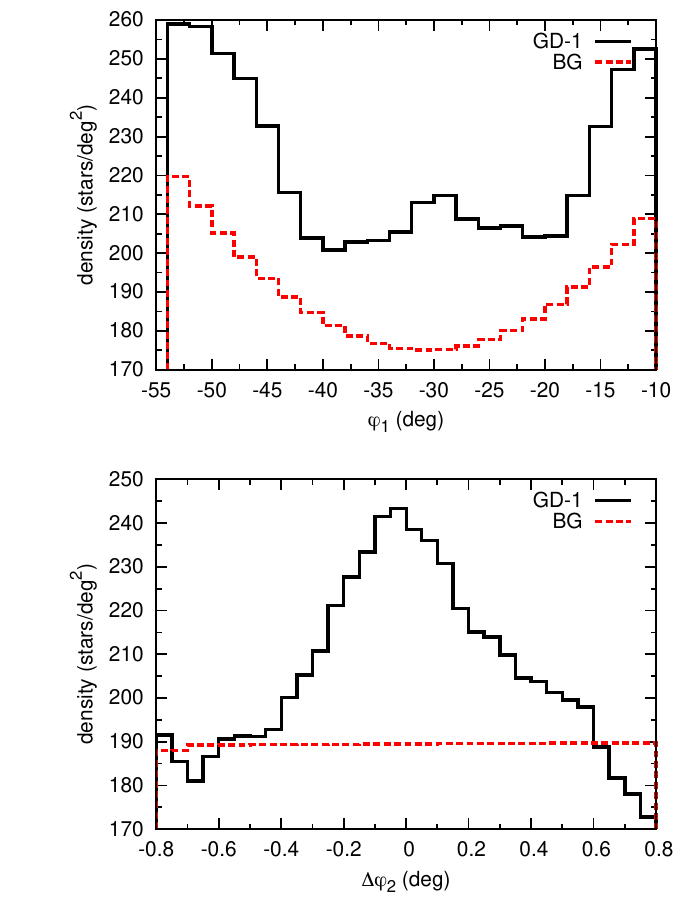}
\includegraphics[angle=0, width=0.495\textwidth]{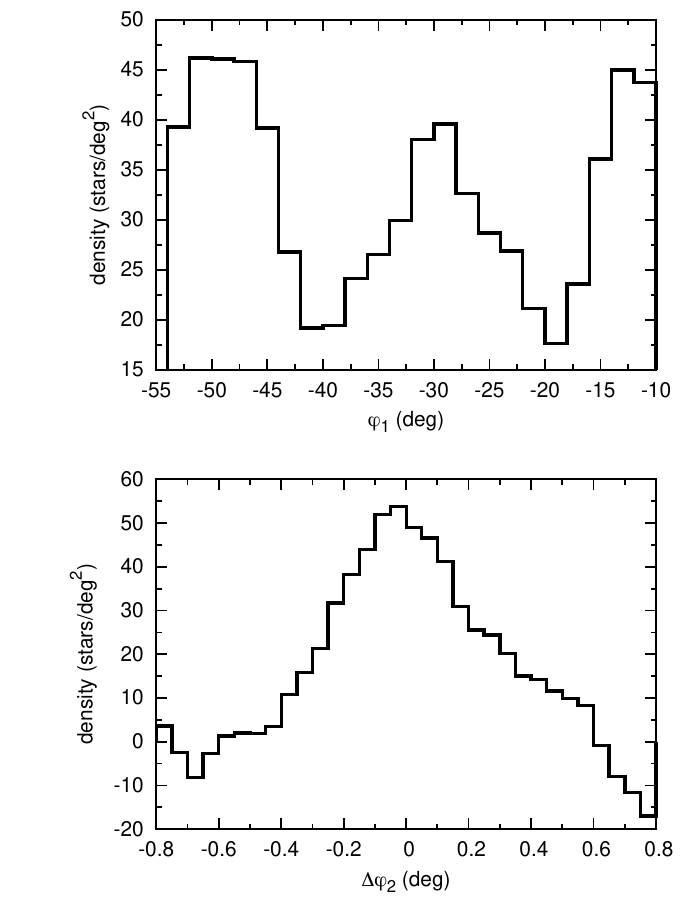}
\caption{Density histograms of GD-1 stars in the matched filter map after stream track correction, as function of $\varphi_{1}$~(top) and $\varphi_{2}$~(bottom). The left panels show the density of the regular and convolved maps along with the density of background sources as fit from the offstream region. The right panels show the density after subtracting the background counts. \label{GD1_denshist}}
\end{figure*}

The normalised density map in the botton panel of Figure~\ref{GD1_MFmap} shows that the stream is thin for $\varphi_{1}>$-40 deg, with a wider section around $\varphi_{1}$=-20 deg. On the left side of the footprint, the stream is much wider and shows clear deviations from the stream track. Most strikingly, at $\varphi_{1}$=-45 deg we see an underdensity centered around $\varphi_{2}$=0 with overdense segment at both higher and lower $\varphi_{2}$. The two stream segments appear to wrap around the central under density, giving the appearance of tidal tails coming off a completely destroyed progenitor. However, we are not claiming to have found the progenitor, but merely commenting on the appearance of this striking feature. We note that perturbations from subhaloes can also produce similar oscillations in the stream track \citep{Erkal15}. The central depletion is clearly visible in the top panel of Figure~\ref{GD1_MFmap}, and consists of an $\approx$20\% drop in stream density. A larger coverage of the GD-1 stream on the low $\varphi_{1}$ side is needed to unambiguously determine the nature of this feature. 

The density of the GD-1 stream can be studied in more detail by constructing density histogram along and perpendicular to the stream track. To correct for the presence of foreground and background contaminants, we construct a model of contaminant sources by fitting a polynomial function to the matched filter densities in the offstream region (0.6$<|\varphi_{2}|<$1.0) where the contaminants dominate. We adopt a second order polynomial in $\varphi_{1}$ and first order in $\varphi_{2}$ to allow for large scale variation as function of Galactic latitude (which lines up largely with $\varphi_{1}$) while the limited coverage of $\varphi_{2}$ only allows a straight line fit. 

Figure~\ref{GD1_denshist} shows the 1D density histograms of GD-1 stars, as function of $\varphi_{1}$~(top) and $\varphi_{2}$~(bottom). The left panels show the density as obtained in Figure~\ref{GD1_MFmap}, while the right panels show the density after correcting for contamination from the Milky Way. The density as function of $\varphi_{2}$ clearly peaks around zero and falls off rapidly to background levels. At positive $\varphi_{2}$, the density is higher than on the negative side, which appears to be due to residual patches of contamination at $\varphi_{2}$=0.5. The density can be fit using a Gaussian profile with a width of 0.16 deg, showing that the GD-1 stream is very narrow across the studied 45 degrees.

The background corrected stream density in Figure~\ref{GD1_denshist} shows GD-1 has a density of $\approx$40 stars/deg$^{2}$ across our range of studied $\varphi_{1}$, punctuated by two large under densities at $\varphi_{1}$=-40 and $\varphi_{1}$=-20 deg. These two density drops are significant and clearly visible in the 2D spatial density maps of Figure~\ref{GD1_MFmap}. The dip at $\varphi_{1}$=-20 deg corresponds to a disturbance in Figure~\ref{GD1_MFmap} where the stream fans out and appears discontinuous. This could be an indication of tidal effects due to a close MW passage or a flyby of a subhalo. Given the distance and orbit of GD-1, an interaction with a giant molecular cloud is unlikely~\citep{Amorisco16}. The under-density at $\varphi_{1}$=-40 deg is wider and deeper and seems linked to the central density depletion at $\varphi_{1}\approx$-45 deg. The two features are not entirely coincident in the 2D and 1D maps due to the added density of the outlying wiggles at $\varphi_{1}$=-45 deg. Interestingly, the presence of underdensities along with stream track variations in a generic prediction of the signature that subhalo perturbations produce in a stream \citep{Carlberg12, Erkal15, Erkal15b}. However, additional modelling is needed to determine if these features are due to an external perturbation or the secular disruption of a progenitor.

The density of foreground and background sources (in red) shows a higher value at either end of our footprint. Figure~\ref{GD1_spatial} shows that the left part of our footprint corresponds to the lowest Galactic latitudes. Qualitatively, the behaviour of the contaminants could be due to an increasing number of MW halo stars for higher $\varphi_{1}$ (i.e. greater Galactic latitude), and an increasing number of thick and thin disk stars at the low latitude end of our footprint. While the density of thin and thick stars is typically much higher than that of the halo, our matched filter preferably leads to halo contamination, which may lead to a greater importance of the halo contribution. However, given the importance of the background model for stream density fluctuations, and the small spatial area used to compute the background density, we use data from the SDSS to check its validity.

\begin{figure}
\centering
\includegraphics[angle=0, width=0.49\textwidth]{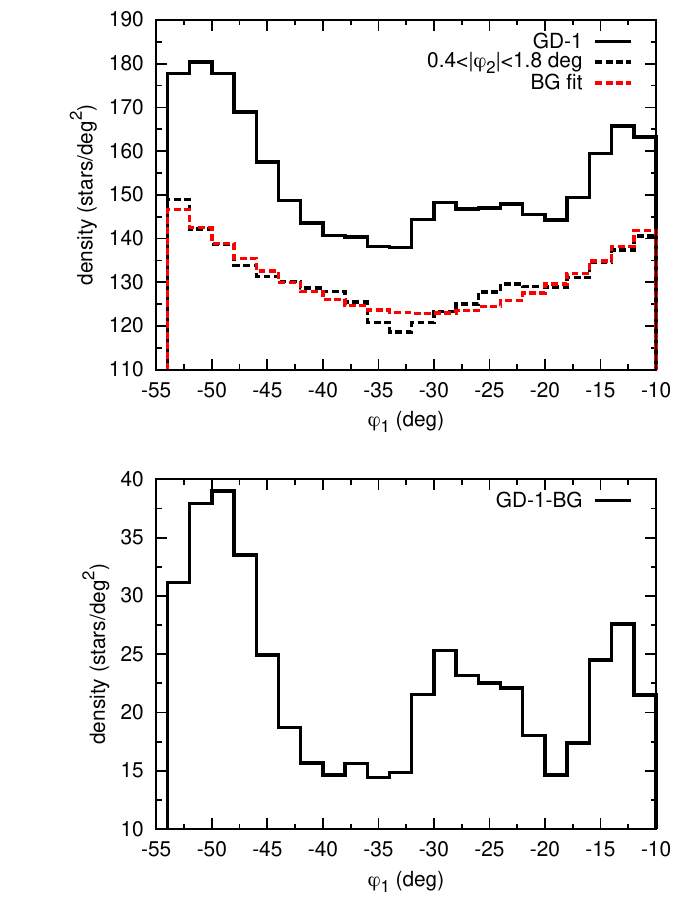}
\caption{\textbf{Top:} Histograms showing the density of GD-1 stars in SDSS after stream track correction, as function of $\varphi_{1}$ for the stream track selection~(solid black line), the outer regions satisfying 0.4$<\varphi_{2}<$1.8 deg~(dashed black line) and the 2D background counts fit. \textbf{Bottom:} The density of GD-1 stars in SDSS as function of $\varphi_{1}$ after subtracting the best-fit background model.\label{GD1_denshist_SDSS}}
\end{figure}

We filter the SDSS data using the exact same procedure and parameters as used to produce Figure~\ref{GD1_denshist} and determine the density of contamination across a much wider area of $\varphi_{2}$. The top panel of Figure~\ref{GD1_denshist_SDSS} shows that the distribution of contaminant sources is remarkably similar in SDSS, along with the over-density of contaminants on either end of the footprint. Therefore, we can rule out effects due to peculiarities of our data from the MW decontamination. Furthermore, the stream density of SDSS sources displays similar under-densities as in our deeper data, albeit with reduced contrast. This gives greater confidence to the validity of our density perturbations.

\section{Conclusions and discussion}\label{conclusions}
In this work, we have utilised deep photometric CFHT data to study the distance, morphology and density of stars in the GD-1 stream. Our deep CMDs of the stream region~(see Figure~\ref{GD1_CMD_trackdistcorr}) show that the data recovers the lower main sequence of the GD-1 stream with unprecedented quality. The photometric error and spatial density distribution (see Figures~\ref{GD1_errs} and~\ref{GD1_spatial}) show that our data sample is homogeneous without obvious depth issues, essential for tracing density variations across the footprint.

The GD-1 stream easily shows up in our CMDs (Figure~\ref{GD1_CMDs}) as a thin main sequence which clearly separates from the MW halo stars due to the data quality. The narrow width of the faint main sequence feature rules out a composite stellar population in GD-1, further indicating that the stream was likely formed from the stripping of a GC, instead of a dwarf galaxy.

Similar to the analysis of~\citet{Koposov10}, we use the data to determine the distance to different parts of the stream in Section~\ref{trackdist}. We find the stream occupies a distance between 8 and 10 kpc in our studied footprint, in good agreement but slightly farther away than~\citet{Koposov10} results. 

\begin{figure}
\centering
\includegraphics[angle=0, width=0.49\textwidth]{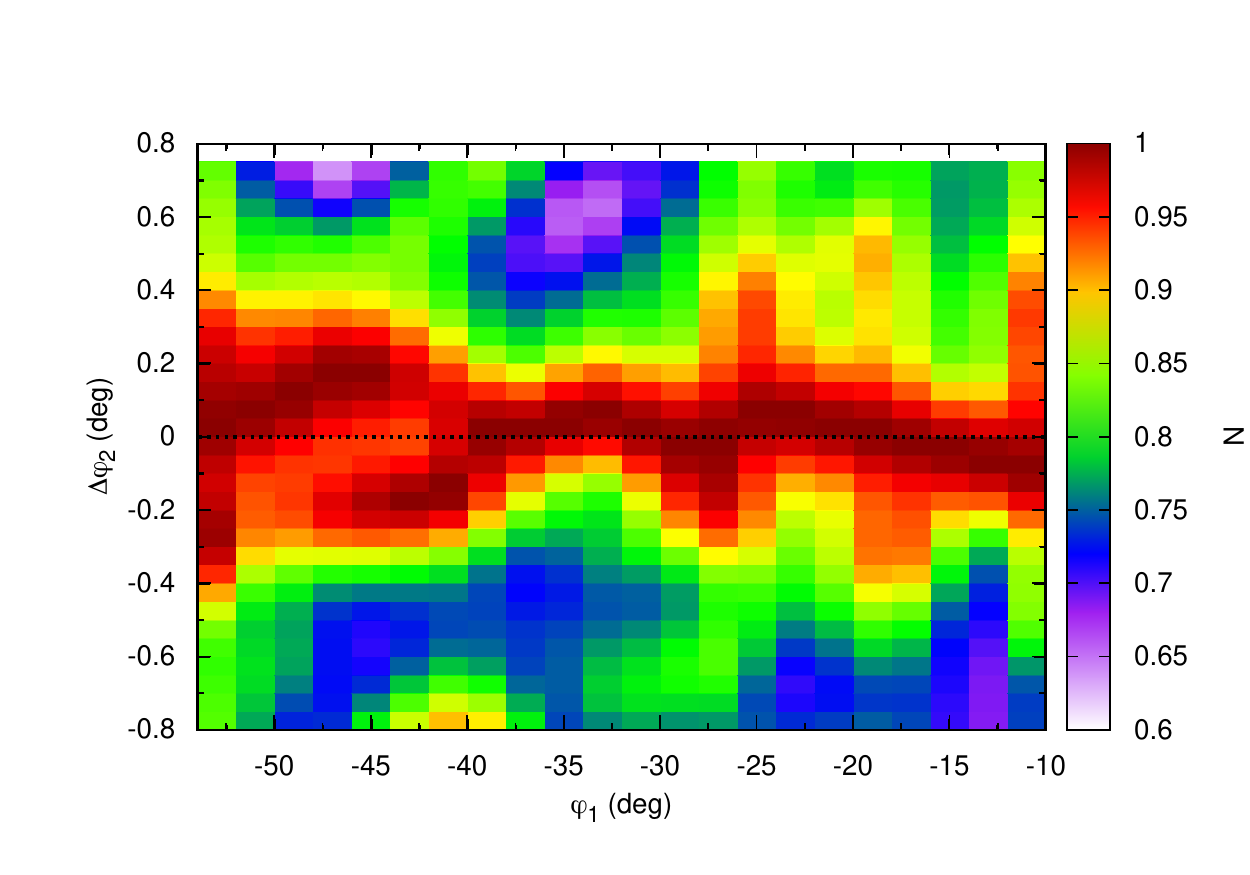}
\caption{Matched filter map of the GD-1 stream from SDSS data, after stream track correction. The same distances, track offsets and CMD selection as in Figure~\ref{GD1_MFmap} are adopted. \label{GD1_MFmap_SDSS}}
\end{figure}

The distance determination makes use of the lower main sequence, thus avoiding degeneracies between distance and age/metallicity. Nonetheless, the lower main sequence in sensitive to binary stars, and a larger binary fraction for GD-1 might therefore induce a distance bias. We have also determined the best-fit stream track by making use of a matched filter approach, following the technique described in~\citet{Erkal17}. The recovered stream track is in agreement with previous results within the uncertainties but tracers the stream density in more detail. The stream track in Figure~\ref{GD1_trackfit} displays some variation on small scales, which could give to rise to artificial wiggles in the track corrected densities. Therefore, we produce a smoothed track using a simple boxcar filter, which is devoid of small scale wiggles. Given the importance of a robust stream track, performing an orbit fitting to stream would result in a smoother and more physically meaningful track. We plan to conduct a full non-parametric fit to the stream data in a forthcoming paper.

The stream track corrected matched filter maps of Figure~\ref{GD1_MFmap} show clear density variations across the 45 deg of GD-1 mapped using CFHT/Megacam. Column normalised maps also show several clear deviations from the track and fanning out of the stream, indicating that GD-1 may have suffered disturbances due to interactions with the MW or other sub-halos. Comparison of the density variations to results from realistic GD-1 simulations in the presence of flybys will be conducted in forthcoming work. Most notably, a clear under-density is seen in the middle of the stream track at $\varphi_{1}$=-45 deg surrounded by overdense stream segments on either side. This location is a promising candidate for the illusive missing progenitor of the GD-1 stream, for further photometric or spectroscopic follow-up. To test for the effects of peculiarities in our survey data, we determine a matched filter map of shallower SDSS data using the same set-up as Figure~\ref{GD1_MFmap}. The column normalised map in Figure~\ref{GD1_MFmap_SDSS} displays similar features, including the central under-density at $\varphi_{1}$=-45 deg. This rules out the possibility of weather, depth variation or tiling strategy being responsible for the observed features discussed in Section~\ref{GD1density}.

\begin{figure}
\centering
\includegraphics[angle=0, width=0.49\textwidth]{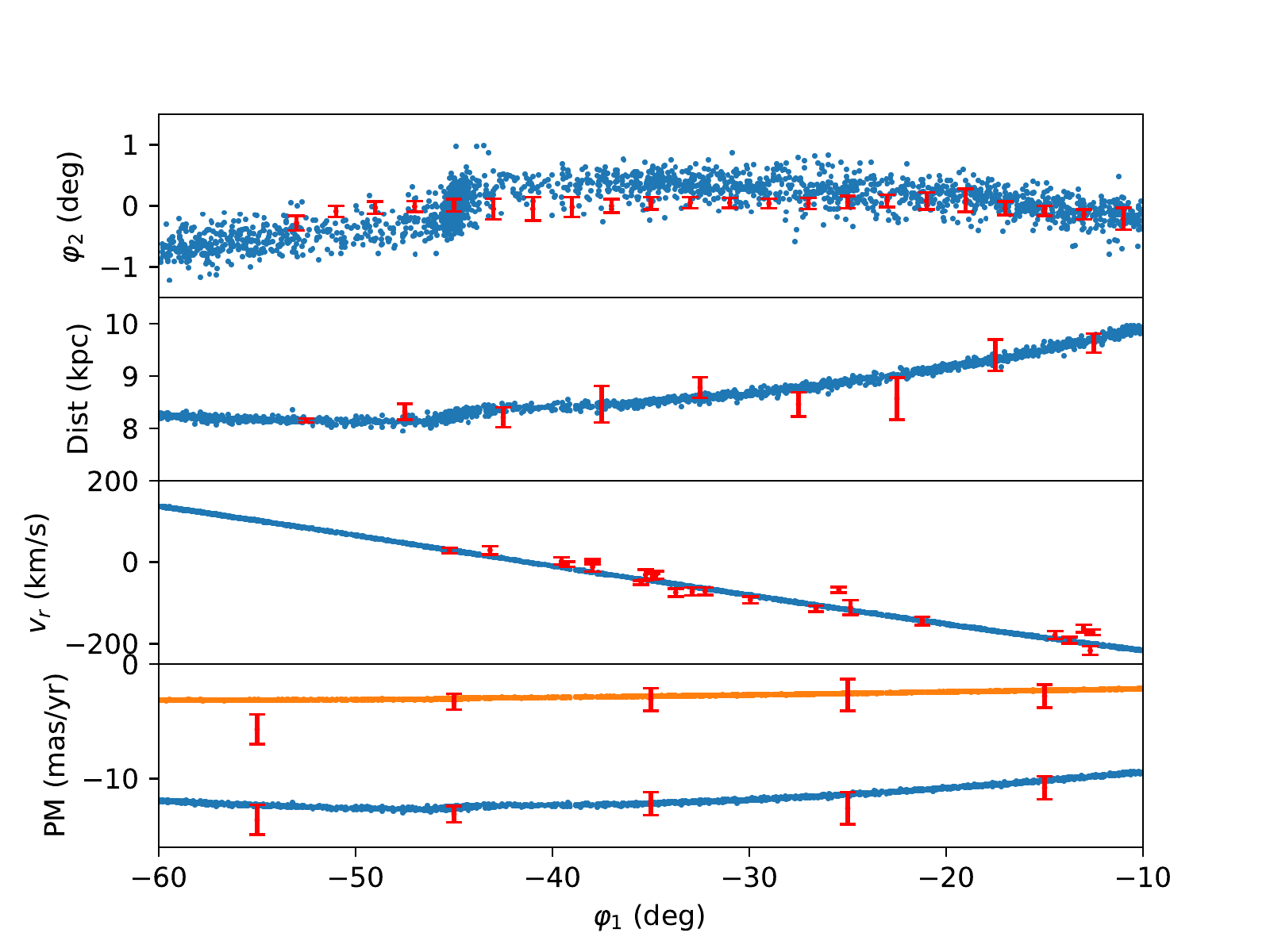}
\caption{Simulation of a GD1-like system dissolving in a realistic MW potential. In this simulation, the progenitor is placed at $\varphi_{1}$=-45 deg at the present to assess the stream track variations resulting from a dissolving progenitor. Different panels show the simulated stream particles in comparison to our stream track (top panel) and distance results (second panel) as well literature radial velocity (third panel) and proper motion measurements (fourth panel) \protect\citep{Koposov10}. Note that in the bottom panel the blue points correspond to $\mu_{\phi 1} \cos \phi_2$ while the orange points correspond to $\mu_{\phi 2}$.  \label{GD1_sim}}
\end{figure}

Projected 1D density histograms (see Figure~\ref{GD1_denshist}) show two large under-densities, one of which is located close to central under-density at $\varphi_{1}$=-45 deg, while the other large dip coincides with a clear stream fanning feature at $\varphi_{1}$=-20 deg. To produce MW decontaminated density histograms, obtaining a robust model of the foreground and background sources is crucial to constrain the underlying stream density. In this work, we have utilised the off-stream region at 0.6$<|\varphi_{2}|<$1.0 to determine the contaminant density and its variation across $\varphi_{1}$. The overall shape of the contaminant source distribution is similar to what is found using data from SDSS (see Fig~\ref{GD1_denshist_SDSS}), ruling out effects due to weather, depth or observational biases on the contaminant density. Nonetheless, obtaining a wider sampling around GD-1 would be recommended when determining the significance of the detected under-densities. 

To assess whether the extent of stream variations and wiggles is indicative of stream disturbance, we have conducted a numerical simulation of a GD1-like system disrupting in a realistic MW potential. The simulation is done with the modified Lagrange Cloud stripping approach of \cite{Gibbons14} to allow us to rapidly sample many progenitor disruptions and get a best fit to the data. The progenitor is modeled as a Plummer sphere with a mass of $10^4 M_\odot$ and a scale radius of 5 pc (which gives roughly the correct stream width on the sky) and is evolved in a realistic Milky Way potential, \texttt{MWPotential2014}, from \cite{bovy_galpy}. The progenitor is placed at $\varphi_{1}$=-45 deg at the present day and then rewound for 3 Gyr, after which it is evolved to the present while tidally disrupting. Figure~\ref{GD1_sim} shows the simulated stream particles in comparison to our stream track and distance results as well literature radial velocity and proper motion measurements \citep{Koposov10}. Note that we simulated approximately 100 disruptions with varying proper motions to get this particular realization of GD01.

The simulated stream particles provide a good fit to the observed properties of GD-1 in both distance and kinematics parameter space. The spatial distribution of the stream displays a large wiggle at $\varphi_{1}$=-45 deg, the location of the progenitor. The qualitative shape of the kink in the stream near the simulated progenitor is quite robust since stars are ejected from the inner and outer Lagrange points \citep[e.g.][]{Kuepper08} so the kink is aligned with the on-sky projection of the radial direction from the Galactic center to the stream. Since the shape of this kink is opposite to the shape seen in the data (see Fig. \ref{GD1_MFmap}), we tentatively conclude that this wiggle is unlikely to be due to a progenitor. Furthermore, beyond the progenitor location the simulated stream track is smooth and continuous, without clear wiggles or gaps of the size as seen in Figure~\ref{GD1_MFmap}. Given the clear presence of track deviations in Figure~\ref{GD1_MFmap} and density variations in Figure~\ref{GD1_distfit}, we therefore conclude that the GD-1 stream may have been disturbed by interactions with the Milky Way disk or other sub-halos. To infer the origin of the wiggles and density depletions, a more detailed comparison with numerical simulations including realistic stream disturbances is needed. 

Given the proximity of the GD-1 stream at 10 kpc, stars on the upper main sequence and turn-off will be present in the second data release of the {\it Gaia} satellite~\citep{GAIAmain1}, which will lead to a much better selection of stream members. The accurate proper motions from {\it Gaia} complemented by the all-sky spectroscopy from WEAVE and DESI will shed new light on the origin of density variations and the location of the missing GD-1 progenitor.

\section*{Acknowledgements}
The authors have enjoyed many a conversation with the members of the
Cambridge Streams Club. T.d.B. acknowledges support from the European Research Council (ERC StG-335936)

Based on observations obtained at the Canada-France-Hawaii Telescope (CFHT) which is operated by the National Research Council of Canada, the Institut National des Sciences de l'Univers of the Centre National de la Recherche Scientifique of France, and the University of Hawaii.

Funding for SDSS-III has been provided by the Alfred P. Sloan
Foundation, the Participating Institutions, the National Science
Foundation, and the U.S. Department of Energy Office of Science. The
SDSS-III web site is http://www.sdss3.org/.

SDSS-III is managed by the Astrophysical Research Consortium for the
Participating Institutions of the SDSS-III Collaboration including the
University of Arizona, the Brazilian Participation Group, Brookhaven
National Laboratory, Carnegie Mellon University, University of
Florida, the French Participation Group, the German Participation
Group, Harvard University, the Instituto de Astrofisica de Canarias,
the Michigan State/Notre Dame/JINA Participation Group, Johns Hopkins
University, Lawrence Berkeley National Laboratory, Max Planck
Institute for Astrophysics, Max Planck Institute for Extraterrestrial
Physics, New Mexico State University, New York University, Ohio State
University, Pennsylvania State University, University of Portsmouth,
Princeton University, the Spanish Participation Group, University of
Tokyo, University of Utah, Vanderbilt University, University of
Virginia, University of Washington, and Yale University.

\bibliographystyle{mn2e_fixed}
\bibliography{Bibliography}

\label{lastpage}

\end{document}